# Uncertainty Modeling in Ultrasound Image Segmentation for Precise Fetal Biometric Measurements

Shuge Lei
University of South Carolina

**Abstract**
Medical image segmentation, particularly in the context of ultrasound data, is a crucial aspect of computer vision and medical imaging. This paper delves into the complexities of uncertainty in the segmentation process, focusing on fetal head and femur ultrasound images. The proposed methodology involves extracting target contours and exploring techniques for precise parameter measurement. Uncertainty modeling methods are employed to enhance the training and testing processes of the segmentation network. The study reveals that the average absolute error in fetal head circumference measurement is 8.0833mm, with a relative error of 4.7347%. Similarly, the average absolute error in fetal femur measurement is 2.6163mm, with a relative error of 6.3336%. Uncertainty modeling experiments employing Test-Time Augmentation (TTA) demonstrate effective interpretability of data uncertainty on both datasets. This suggests that incorporating data uncertainty based on the TTA method can support clinical practitioners in making informed decisions and obtaining more reliable measurement results in practical clinical applications. The paper contributes to the advancement of ultrasound image segmentation, addressing critical challenges and improving the reliability of biometric measurements.

**Key words:** Ultrasound imaging; Uncertainty Estimation; Image segmentation.

## A. Introduction

Medical image segmentation stands as a critical component within the domain of computer vision and medical imaging. The process of partitioning images into distinct regions, delineating boundaries, and identifying objects or structures plays an integral role in numerous applications, spanning from object detection to diagnostic tools in healthcare. Within this domain, ultrasound imaging stands as a valuable modality offering real-time, non-invasive visualization of internal anatomy. Despite its utility, ultrasound imaging presents unique challenges for accurate and precise segmentation due to inherent image noise, variable image quality, and the subjective nature of acquisition methods.

Segmentation in medical ultrasound images assumes significant importance in clinical diagnostics and treatment planning. However, it remains a challenging task owing to several factors. The inherently noisy nature of ultrasound images, coupled with the ambiguity of organ boundaries and varying tissue textures, poses substantial obstacles to accurate and consistent segmentation. Moreover, the variability in imaging settings, transducer placements, and operator expertise results in differences in image quality and acquisition protocols, further complicating the segmentation process.

The accuracy and interpretability of segmentation outcomes are crucial in ultrasound imaging for facilitating early disease detection, treatment monitoring, and surgical planning. Ensuring precise delineation of anatomical structures and pathological regions is essential for clinical decision-making. However, due to the complexity and uncertainty inherent in ultrasound data, achieving reliable and accurate segmentation remains a pressing challenge.

This paper explores the intricacies of uncertainty in medical image segmentation, specifically focusing on ultrasound data. To enhance the understanding of uncertainty sources, particularly within ultrasound images, this project proposes a methodology to visualize and mitigate uncertainty in segmentation results. We leverage the PFNet (Positioning and Focus Network) architecture and extend its capabilities to address the unique challenges posed by ultrasound imaging. By investigating uncertainty sources and employing tailored techniques within PFNet, this work seeks to improve the reliability, accuracy, and interpretability of ultrasound image segmentation.

## B. Related work
### A. Segmentation networks

The task of medical image segmentation is a fundamental challenge in the domain of computer vision and medical image analysis. Various classic segmentation networks have been proposed in the literature, each with its unique attributes and applications. Some noteworthy examples include Fully Convolutional Networks (FCN)[1], U-Net Convolutional Networks (U-Net)[2], Semantic Segmentation Net (SegNet)[3], DeepLab[4], and V-Net Convolutional Networks[5]. These networks are adept at achieving pixel-level and multi-class segmentation. The U-Net architecture, for instance, employs an encoder-decoder structure that leverages down-sampling and increased channel capacity to extract deeper semantic information from images. This design facilitates the propagation of contextual information to higher resolutions and reduces information loss during gradient descent through skip connections.

This research build upon the PFNet (Positioning and Focus Network) segmentation network to address challenges in ultrasound medical image segmentation[6]. PFNet, originally designed for camouflaged object segmentation (COS), specializes in fine-grained segmentation tasks, particularly in scenarios where the target closely resembles the background. The model consists of two critical components: the Positioning Module (PM) and the Focus Module (FM). The Positioning Module is formed by concatenating Channel Attention (CA) and Spatial Attention (SA) modules, while the Focus Module fine-tunes the foreground boundaries obtained from the Positioning Module, addressing false negatives and false positives. Due to variations in fetal developmental stages and differences in medical imaging techniques, the scales of foreground objects vary, necessitating the consideration of multiple resolutions. PFNet adeptly transits from coarse-grained to fine-grained segmentation, focusing on refining initial segmentation results, particularly by addressing ambiguous regions and integrating low-resolution information into high-resolution layers, with a focus on precise contour boundary delineation.

This paper extends PFNet's capabilities by embedding image knowledge. In the post-processing phase, contour non-maximum suppression is applied to mitigate false positives through the enforcement of boundary conditions, resulting in noise reduction and the extraction of the maximum foreground contours. This work focuses on applying the PFNet to address the specific challenges associated with medical image segmentation on ultrasound data.

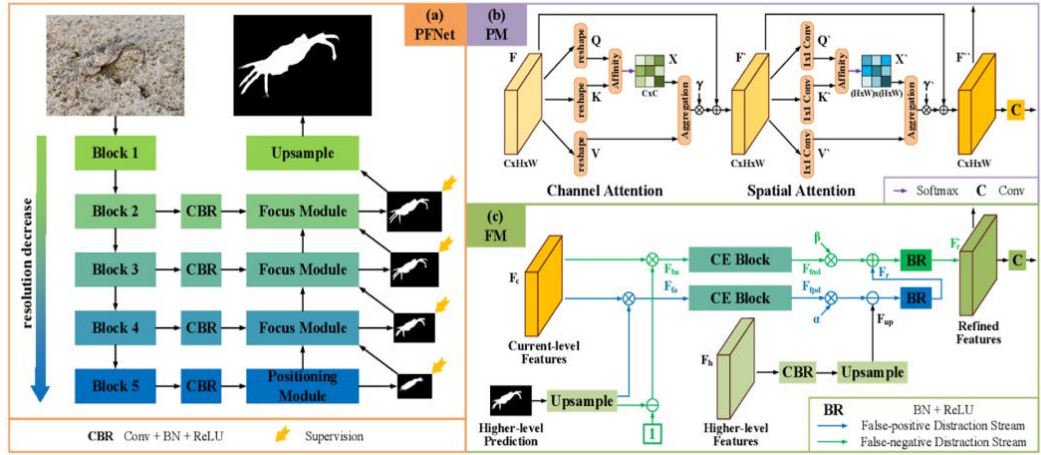

Figure 1: PFNet Network Architecture [6]

### B. Uncertainty Measurement

The model (epistemic) uncertainty and data (aleatoric) uncertainty can be visually depicted as in Figure 2. For binary classification tasks (where "circles" and "crosses" in the figure represent the sample spaces of two classes), the classification boundary is notably ambiguous. The threshold of 0.5 is often employed for classification; yet a classifier based on the 0.5 threshold is not necessarily well-calibrated. Samples in the overlapping region exhibit data (aleatoric) uncertainty, the space domain that is not encompassed within the training space denote model (epistemic)uncertainty.

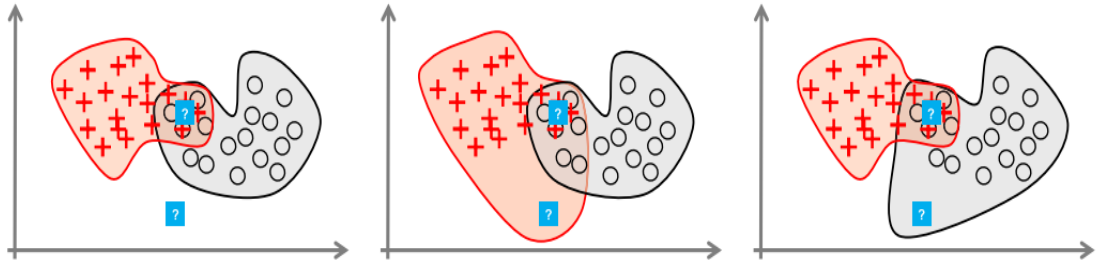

Figure 2: Epistemic and aleatoric uncertainty [7]

### C. Test Time Data Augmentation

From a data-centric perspective, the Test Time Augmentation (TTA) method is employed to expand the sample space with augmented data, thus broadening the model's epistemic range and reducing model epistemic uncertainty for improved neural network calibration. Regarding data augmentation, it's crucial to consider the inherent characteristics of medical images, such as mimicking variations in images acquired by different devices through changes in color and saturation, and simulating variations in physician operation techniques through adjustments in brightness, contrast, translation, and scaling. When the same image got transformed T times via TTA, the outcome is no longer a point estimate but a probability distribution of predictions for the same sample. This is conditioned on that the labels for the same image remain unchanged after variations in lighting, rotation, and translation.

While the blurriness and downsampling have relatively minor impact to the object recognition task, they have significant impacts on image super-resolution. Therefore, this paper only considers spatial transformations and inherent noise. The image augmentation process can be viewed as transforming the original image through a certain transformation model to obtain a new image as in (1).

$$X = \Gamma_\beta(X_0) + e \quad (1)$$

In this context, $X_0$ represents a latent variable, denoting a low-level image at a specific position and orientation. $\Gamma$ is a transformation applied on $X_0$, and $\beta$ represents the hyperparameters of the transformations. $e$ denotes noise, and $X$ represents the transformed images. While transformations can occur in space, intensity, or feature space, this paper only considers reversible spatial transformations, which include flipping, scaling, rotation, and translation. $\Gamma_\beta^{-1}$ represents the inverse transformation of $\Gamma_\beta$.

$$X_0 = \Gamma_\beta^{-1}(X - e) \quad (2)$$

It is typically assumed that the transformation parameters and noise follow some prior distribution. For example, the rotation angle, denoted as $r$, can be modeled as a uniform prior distribution $r \sim U(0, 2\pi)$. Image noise can be modeled as a Gaussian distribution $e \sim N(\mu, \sigma)$, where $\mu$ and $\sigma$ represent the mean and variance of the noise. $p(\beta)$ and $p(e)$ represent the prior distributions of $\beta$ and $e$, resulting in $\beta \sim p(\beta)$ and $e \sim p(e)$. Assuming that $Y$ and $Y_0$ are the labels for $X$ and $X_0$ regarding image classification and segmentation problems. $Y$ and $Y_0$ are discrete labels, and they should be equal before and after spatial transformations, i.e., $Y = \Gamma_\beta(Y_0)$.

The testing images are fed into a neural network to obtain predictions:

$$Y' = f(\theta, X) \quad (3)$$

$Y'$ represents the network's predictions, which, in the context of classification and segmentation problems, denote the probability distribution of the network's final layer output. Given that X is but one of numerous potential observations, inferring Y directly from X may yield biased results, necessitating the use of latent variables $X_0$ to infer Y:

$$Y = \Gamma_\beta(Y_0) = \Gamma_\beta\big(f(\theta, X_0)\big) = \Gamma_\beta(f(\theta, \Gamma_\beta^{-1}(X - e))) \quad (4)$$

Here, $\beta$ and $e$ are unknown. $Y$ is not an exact prediction but rather a distribution of Y, considering the distribution of Y given $\beta$ and $e$:

$$p(Y|X) = p\left(\Gamma_\beta\left(f\left(\theta, \Gamma_\beta^{-1}(X - e)\right)\right)\right), \quad \beta \sim p(\beta), e \sim p(e) \quad (5)$$

For continuous variables, Y can be estimated using the expectation:

$$E(Y|X) = \int y p(y|X) dy$$

$$= \int \int_{\beta \sim p(\beta), e \sim p(e)} \Gamma_\beta\left(f\left(\theta, \Gamma_\beta^{-1}(X - e)\right)\right) p(\beta) p(e) d\beta de \quad (6)$$

Equation (6) involves significant computational complexity, where $\beta$ and e are continuous values, and $p(\beta)$ represents the joint distribution of different transformations. This can be estimated through a Monte Carlo approach. Using N to denote the number of random samples, the prediction for the $nth$ sample is:

$$y_n = \Gamma_{\beta_n}\left(f\left(\theta, \Gamma_{\beta_n}^{-1}(X - e_n)\right)\right), \beta_n \sim p(\beta), e_n \sim p(e) \quad (7)$$

To get $y_n$, hidden variable $X_0$ needs to be calculated first via independently sampling η and ε from their respective prior distributions $p(\beta)$ and $p(e)$. These samples, based on the current sampling, are then input into the network. Thus, the estimate of $E(Y|X)$ can be obtained as (8).

$$\hat{Y} = E(Y|X) \approx \frac{1}{N}\sum_{n=1}^{N} y_n \quad (8)$$

For classification and segmentation problems, the final prediction results are obtained through a majority vote, which represents the maximum likelihood estimate:

$$\hat{Y} = argmax\, p(y|X) \approx Mode(y) \quad (9)$$

**D. Monte Carlo Dropout**

MC Dropout is one of the extensible techniques in Bayesian neural networks, initially widely adopted by researchers as a regularization method to enhance the network's generalization ability. In this paper, the MC Dropout is used to approximate the Bayesian modeling process as the method proposed by Yarin Gal et al. [8]. From a modeling perspective, by stochastically sampling inactive neurons, this method introduces randomness to the model's parameters, enhancing the model's capacity to fit the feature space, thus expanding the model's epestemic scope. This makes the network's output predictions closer to the true probability distribution, resulting in calibrated output probabilities.

This method assumes that $q(\theta)$ is an approximate distribution on the network parameter set $\theta$, with its elements being stochastically set to zero based on a Bernoulli random variable distribution. The approximation of $q(\theta)$ can be achieved by minimizing the posterior distribution given the training set. After training, the predictive distribution can be expressed as:

$$p(Y|X) = \int p(Y|X,\omega)q(\omega)\,d\omega \quad (10)$$

An approximation of $p(y|x)$ can be obtained through Monte Carlo sampling from $y_n = f(\theta_n, X)$, where $\theta_n$ is obtained from $q(\theta)$ through Monte Carlo sampling. For classification problems, the final prediction can be expressed as Equation (11):

$$E[y] \approx \frac{1}{T}\sum_{t}^{T} \hat{y}_t(x) \quad (11)$$

For regression problems, the model's predicted mean and variance can be expressed in the following form:

$$Var_q(y) \approx \tau^{-1}I_D + \frac{1}{T}\sum_{t}^{T} \hat{y}_t(x)^T \hat{y}_t(x) - E[y]^T E[y] \quad (12)$$

In the above equations, $x$ represents the input, $\hat{y}t(x)$ is the corresponding output, $\tau$ is a constant defined by the model structure, and T is the times of the random forward passes during testing.

### III. Proposed Method

Uncertainty Measurement in Segmentation Problems. Image segmentation can be regarded as a pixel-level classification task, where the network generates a probability distribution for each pixel. This paper employs TTA (Test-Time Augmentation) and MC Dropout to measure uncertainty in medical image segmentation. Model (epistemic) uncertainty and data (aleatoric) uncertainty are calculated on a per-pixel basis using equations (14) and (15) for both methods. Model uncertainty (i.e., MI) is obtained by subtracting data uncertainty from the total uncertainty.

In the context of K-class classification tasks, consider the probability distribution of the network output corresponding to input sample $x$, denoted as $p(x)$ and abbreviated as $p$. Here, $p_k$ represents the output probability of the $k_{th}$ category in the vector. Generally, the given prediction $p$ represents a classification distribution, indicating the assigned probabilities for each category prediction of the input sample. Since the prediction is a probability distribution, uncertainty estimation can be directly derived from the prediction.

To estimate the data uncertainty of the prediction, entropy is commonly used as a calculation method. The maximum category probability represents the deterministic outcome of the network output, while entropy describes the information encompassed by all category predictions in the forecast.

$$Maximal\ probability: p_{max} = \max\{p_k\}_{k=1}^{K} \qquad (13)$$

$$Entropy: H(p) = -\sum_{k=1}^{K} p_k \log_2(p_k) \qquad (14)$$

While the output of softmax activation theoretically contains information about the uncertainty of predictions, it is challenging to discern the extent of model uncertainty affecting the final prediction from a single prediction. A single softmax prediction is not a highly reliable method for quantifying uncertainty and does not provide any information about the model's inherent certainty reding a specific output [9].

Approximating posterior distributions $p(\theta|D)$ on model parameters holds promise for obtaining better uncertainty estimates. With the posterior distribution, the softmax output itself becomes a random variable, allowing for the assessment of its variability, i.e., uncertainty. Commonly used measurement methods include Mutual Information (MI), Expected Kullback Leibler (EKL) divergence, and prediction variance[7]. In general, all these measurement methods involve computing the expected divergence between the random softmax output and the expected softmax output.

Among these, MI employs information entropy to calculate model (epistemic) uncertainty, often subtracting the component of data uncertainty from the total uncertainty information to obtain model uncertainty. The calculation formula is as follows:

$$MI(\theta, y | x, D) = H[\hat{p}] - E_{\theta: p(\theta|D)} H[p(y | x, \theta)] \qquad (1)$$

Smith et al. pointed out that when the knowledge of model parameters does not increase the information in the final prediction, Mutual Information (MI) is minimized [9]. Kullback-Leibler divergence measures the divergence between two given probability distributions. Expected Kullback Leibler (EKL) can be used to measure the expected divergence between possible softmax outputs, serving as an interpretation of the uncertainty in the model output[7]. Therefore, it represents a measure of the model's uncertainty.

$$E_{\theta: p(\theta|D)}[KL(\hat{p} \| p)] = E_{\theta: p(\theta|D)}\left[\sum_{k=1}^{K} \hat{p}_i \log\left(\frac{\hat{p}_i}{p_i}\right)\right] \qquad (2)$$

Prediction variance assesses the variance of random softmax outputs:

$$\sigma(p) = E_{\theta: p(\theta|D)}\left[(p - \hat{p})^2\right] \qquad (3)$$

Only Bayesian methods can provide an analytical description of the posterior distribution $p(\theta|D)$. Even with an analytical description, dealing with parameter uncertainty in almost all prediction scenarios is challenging and often requires approximation through Monte Carlo methods. Similarly, ensemble methods involve obtaining the final prediction from *M* neural networks; while at test time, data augmentation methods yield *M* predictions from *M* different augmentations of the original input sample. In all these cases, a set of predictions with a sample size of *M* can be obtained, which can be used to approximate challenging or even undefined potential distributions. With these approximations, metrics defined on MI, EKL, Sigma can be directly applied, but the posterior distribution must be replaced with the average value. For example, the expected output of softmax becomes:

$$\hat{p} \approx \frac{1}{M} \sum_{i=1}^{M} p^i \qquad (4)$$

Then Mutual Information (MI) represents model uncertainty, while $\mathrm{E}_{\theta: P(\theta|D)} H[p(y|x,\theta)]$ represents data uncertainty.

## IV. Experiments and Results

### A. Datasets and metrics

**Datasets.** The experimental data for this study is from the publicly available dataset of fetal thalamic ultrasound standard sections and the private fetal femur ultrasound dataset collected in the laboratory at Peking University Shenzhen Hospital. The private dataset includes: (1) 999 fetal thalamic ultrasound standard sections, covering various stages of fetal development. Each image includes pixel size(mm) and clinical measurements of fetal head circumference(mm), as in Figure 2. (2) 230 fetal femur ultrasound standard sections, each with pixel distances and clinical measurements of fetal femur length, as in Figure 3.

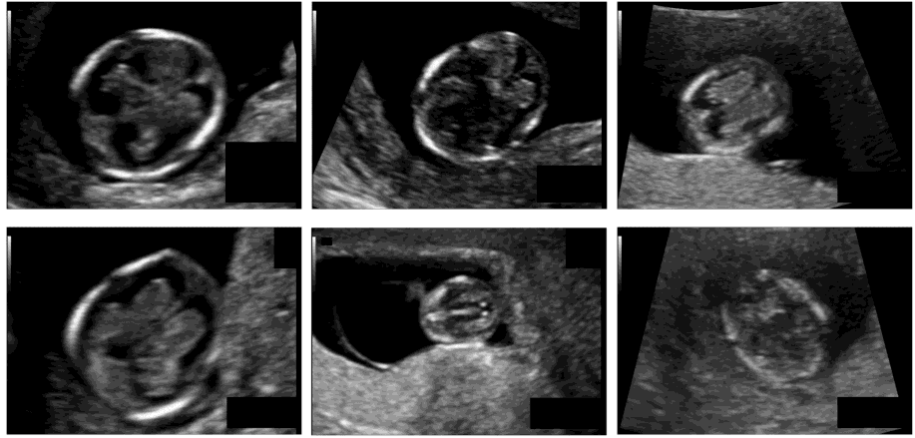

Figure 2: Fetal Head Ultrasound Standard Sections and Head Circumference Dataset

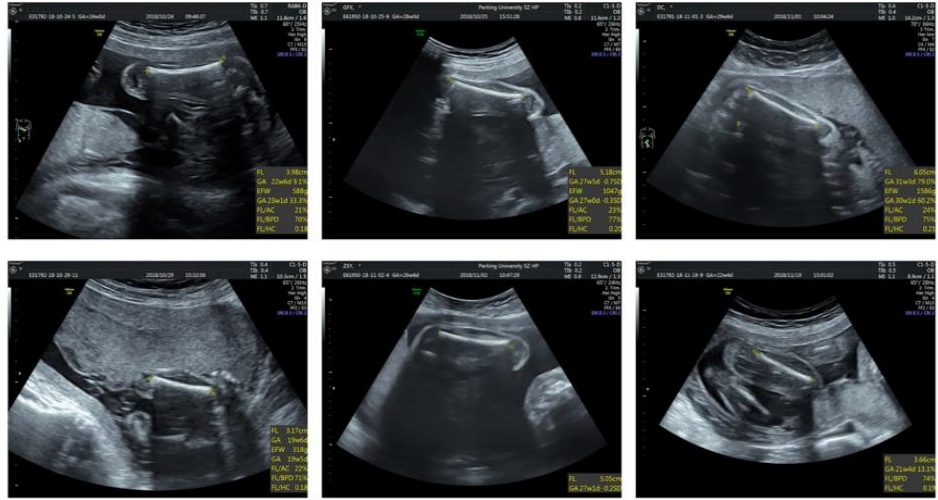

Figure 3: Fetal Femur Ultrasound Standard Sections and Femur Length Dataset

**Metrics.** Intersection over Union (IOU) is used to evaluate segmentation performance and localization accuracy. It's worth noting that the dataset labels (the clinical measurements) in the dataset are based on ellipse annotations, which are clinical approximations of fetal head circumference. However, models trained with coarse labels often produce polygonal predictions rather than ellipses. Thus the IOU calculation in this paper is provided for reference purposes. The IOU calculation method, as shown in Figure 4, is obtained using TP/TP+FP+FN to determine the IOU value between the predictions and the labels.

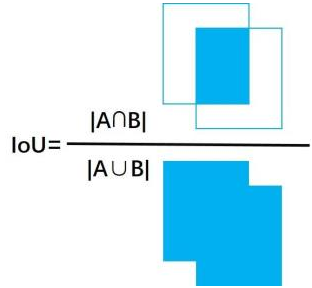

Figure 4: IOU calculation

A relative error metric is used to evaluate the discrepancies between prediction results and clinical measurements, as shown in Equation (10), where $x$ represents the predicting measurement value, and $u$ represents the ground truth.

$$relative\ error = \frac{|x - \mu|}{\mu} \times 100\% \quad (19)$$

Experimental Parameter Settings. The dataset is divided into training, validation, and test subsets at a 6:2:2 ratio. Test-Time Augmentation (TTA) technique involves augmenting data such as horizontal flipping, scale resizing, rotation, brightness, contrast, saturation, and hue adjustments. In the testing phase, the segmented images obtained from predictions undergo inverse transformations including horizontal flipping, scale resizing, and rotation. These transformed outcomes are averaged and subsequently binarized, culminating in the derivation of the definitive binary segmentation image. The process is repeated consistently for a specified number of test iterations, denoted as T=8.

In MC Dropout, dropout mechanisms are employed across the network layers, specifically within the output prediction segments of the three Focus modules. These correspond to the preceding layers of the fully connected layer in a per-pixel classification network. The final convolutional layer maps each pixel point from a high-dimensional channel to a value within the range of [0, 1]. During the testing phase, dropout is activated while the Batch Normalization (BN) layers within the network are deactivated. The augmented identical image is input into the network repeatedly to generate multiple prediction outcomes. Testing iterations are set at T=8, and the ensuing results are aggregated, averaged, and binarized, yielding the final binary segmentation image as illustrated in Figure 5.

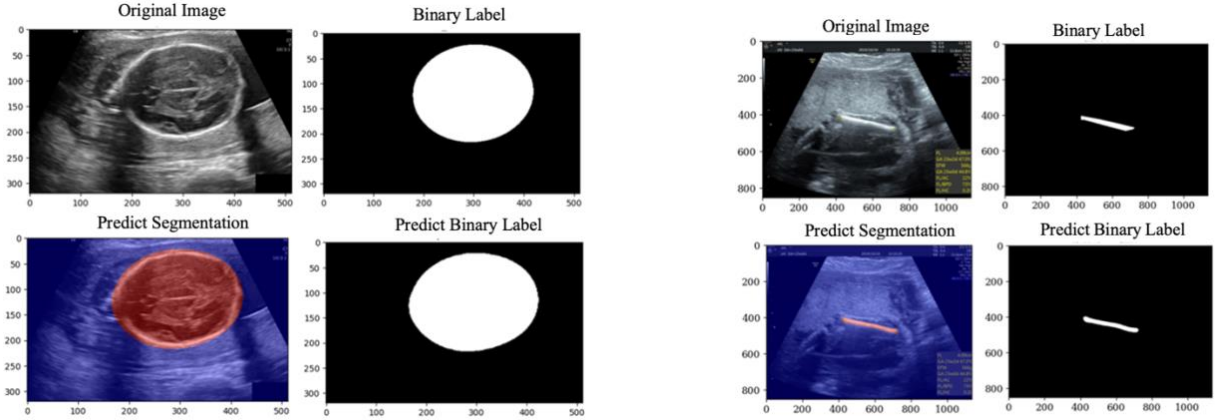

Figure 5: Fetal Ultrasound Original Image, Original Image Label, and Predict Segmented Binary Image

Experimental Parameters for PFNet. Input images are resized to a resolution of 416*416 for training. The encoding network employs a ResNet-50 model pre-trained on ImageNet, while the parameters of other layers are initialized randomly. Stochastic Gradient Descent (SGD) optimizer is used with a weight decay coefficient $5 * 10^{-4}$, and a batch size is set to 16. The learning rate is adjusted through the Poly strategy using the following formula:

$$lr = base\_lr \times (1 - \frac{epoch}{num\_epoch})^{power} \quad (20)$$

$Base\_lr$ is set to 0.001, and the power is configured as 0.9. The images are mapped back to the original size of the input images after being resized to 416*416. Both of these adjustment processes employ bilinear interpolation.

Experimental Environment. The PyTorch toolbox [11] is utilized, and two NVIDIA GeForce RTX 2080Ti GPUs (each with 11GB of memory) are employed for training and testing.

## B. Experimental Results

The Intersection over Union (IOU) between the segmentation results and ground truth (GT) labels is computed for the baseline model, TTA model, and MC Dropout model. The results are presented in Table 1.

Table 1. Segmentation IOU

|  | Baseline | TTA | MC Dropout |
|---|---|---|---|
| Fetal Head IOU | 0.9664 | 0.9690 | 0.9655 |
| Femur IOU | 0.8528 | 0.8349 | 0.8154 |

Fetal Head Circumference Measurement. The segmentation-derived object contour is fitted with an ellipse, and its perimeter is calculated as the measurement of head circumference. This measurement is compared with the ground truth elliptical label to calculate both absolute and relative errors. OpenCV [12] is utilized for ellipse fitting. The predicted mask is the input to OpenCV and the minimum enclosing rotated rectangle of the fitted ellipse is the output. The major and minor axes, and the center coordinates of the fitted ellipse can be obtained from the rectangle. The ellipse's perimeter is calculated using Equation 21:

$$L = 2\pi b + 4(a - b) \qquad (21)$$

'a' and 'b' respectively represent the major and minor axes of the ellipse. In the baseline model, the average absolute error between the ellipse-fitted object contour and the ground truth label is 8.0833 mm, with an average relative error of 4.7347%. Figure 6 presents an example of the segmentation result of the fetal head and the ellipse fitting of its contour.

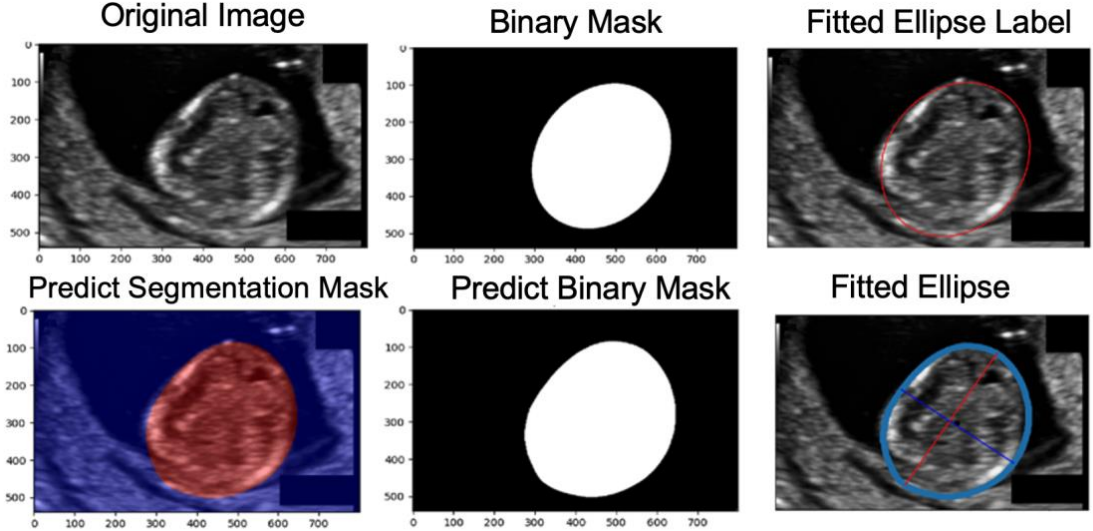

Figure 6: Example of fetal head segmentation results and contour ellipse fitting

Femur Length Measurement. Unlike measuring the circumference, measuring the length involves the actual measurement of the distance between two endpoints in the target area. In this study, the measurement task is simplified as a segmentation task. Based on the obtained object contour, the minimum bounding rectangle is calculated, and the longer side of this rectangle is considered an approximate distance between the two endpoints. Figure 7 shows the network-predicted fetal femur contour (blue curve) and the minimum bounding rectangle (red rectangular box). For clarity, a portion of the image enclosed in the orange box has been magnified. In the baseline model, the average absolute error for femur length measurement is 2.6163 mm, with an average relative error of 6.3336%.

Figure 8 presents the segmentation results of the fetal femur. Since the binary labels are based on polygonal annotations and not fine-grained labels, the segmentation IOU value can only provide a reference for segmentation accuracy and localization accuracy. A high IOU value ensures both segmentation and localization accuracy, while a low IOU value indicates potential issues with localization accuracy, especially when the predicted contour area is significantly smaller than the target contour, resulting in lower segmentation accuracy.

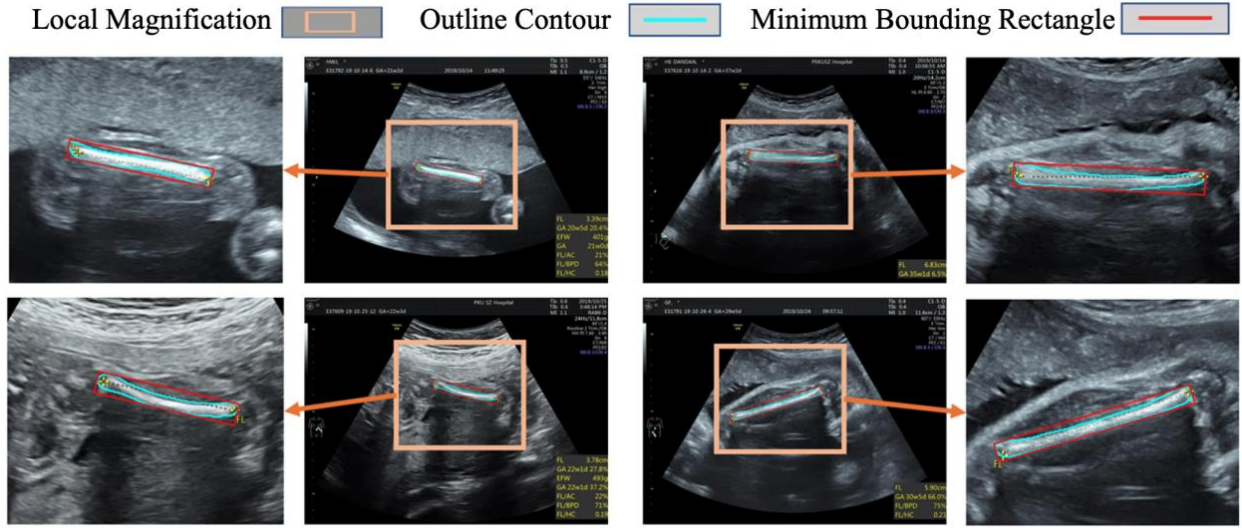

Figure 7: Fetal Femur Predicted Contour (Blue) and Minimum Bounding Rectangle (Red)

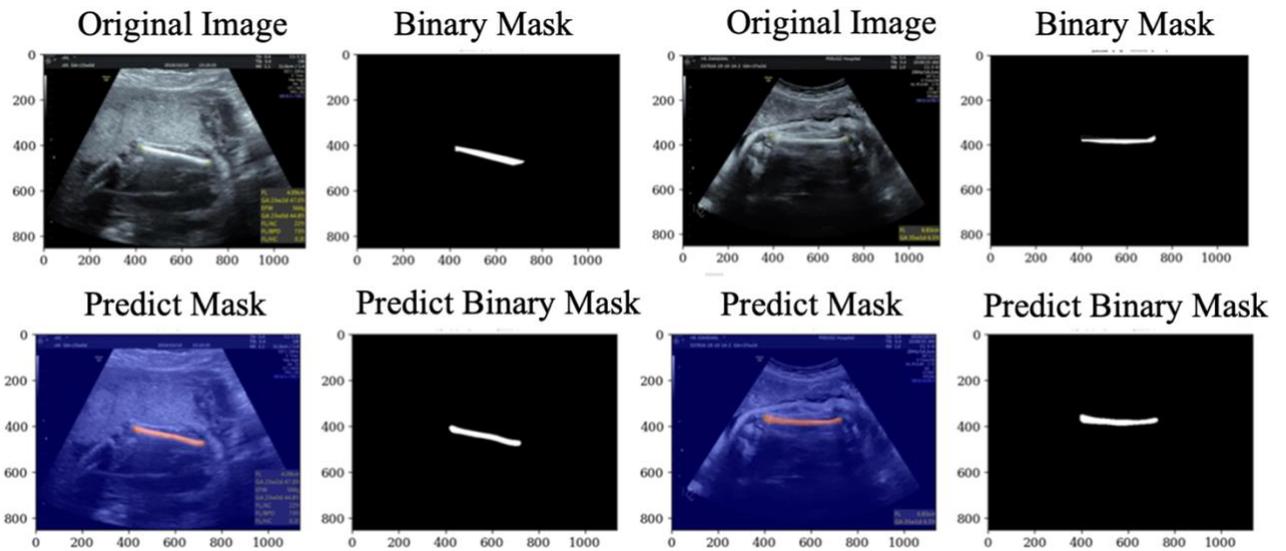

Figure 8: Fetal Femur Segmentation Results

### C. Uncertainty Modeling Experimental Results

Figures 9 and 10 depict alternating rows displaying original images, ground truth (GT) labels, and the final segmented contours obtained by averaging multiple predictions followed by binarization. Even rows display the data uncertainty and model uncertainty predicted by the model for the current sample, with colors indicating the degree of uncertainty (larger values correspond to greater uncertainty). Here are some findings from the color values in the computed results: (1) The TTA method primarily models data uncertainty, while MC Dropout focuses primarily on model uncertainty. (2) Model uncertainty obtained from MC Dropout is mainly concentrated on the edges of the segmented foreground, with pixels further away from the boundary exhibiting very low uncertainty. (3) Data uncertainty obtained from TTA not only shows higher uncertainty at the edges but also exhibits higher uncertainty in challenging areas, as indicated by the white arrows. In these regions, TTA models produce inaccurate segmentations, where the segmentation boundary lies outside or inside the GT boundary. This corresponds to the high values in the same region of the arbitrary uncertainty map.

Both the epistemic uncertainty maps in Figure 9 and Figure 10 display relatively low uncertainty in areas with segmentation errors. This can result in the network being "overconfident" in its incorrect results. Conversely, data uncertainty obtained through TTA shows higher values in error-prone areas (as indicated by the white arrows). This comparison indicates that data uncertainty is better at identifying errors in the segmentation of non-boundary pixels. For these pixels, the segmentation output is more influenced by different input transformations (aleatoric) than changes in model parameters (epistemic).

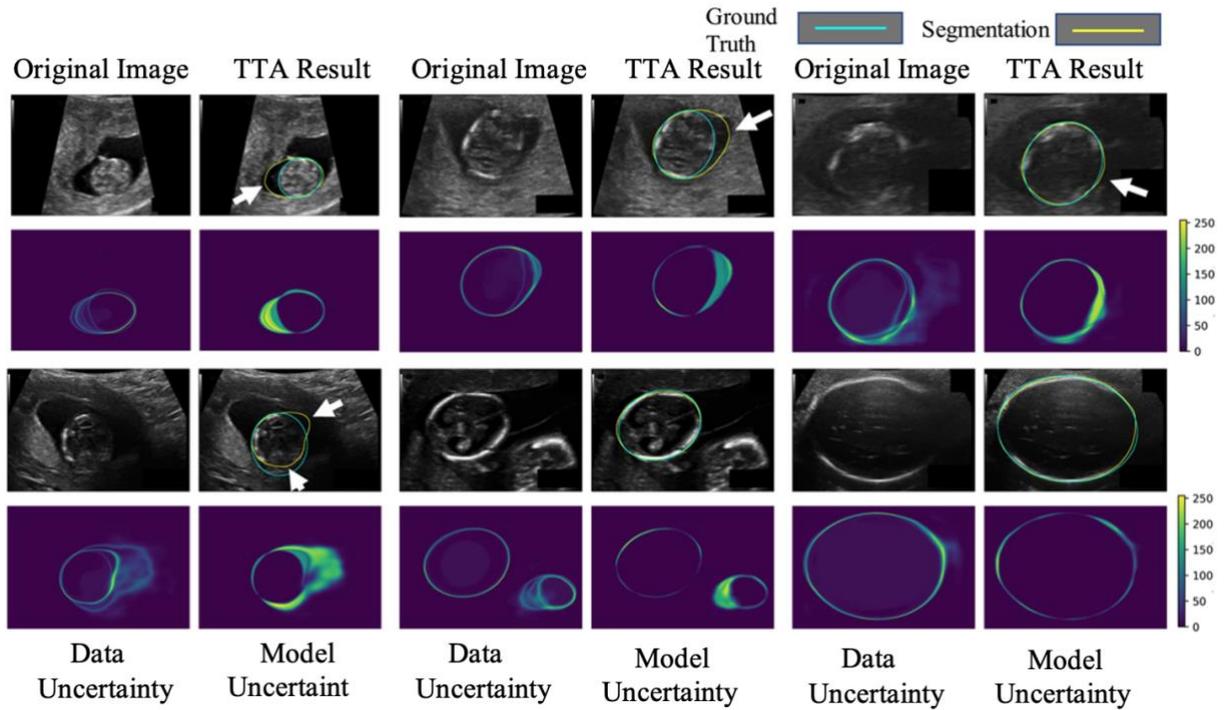

Figure 9: Quantitative Results of Different Types of Uncertainty in Fetal Brain under the TTA Method

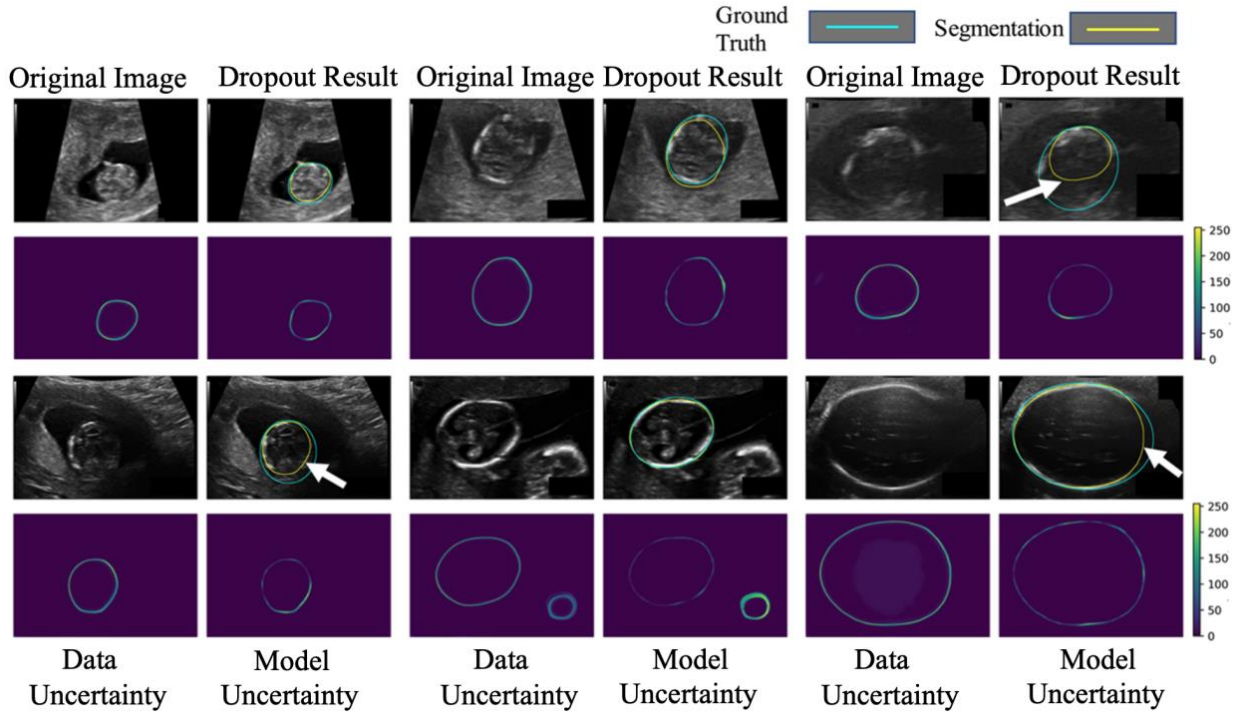

Figure 10: Quantitative results of different types of uncertainty in fetal brain under the Dropout method

**Uncertainty Pattern Exploratory**. To explore how the uncertainty is related with the error rates, the distribution of error rates under different uncertainty values is showed in the form of histograms as Figure 11. For all pixels contained in each uncertainty map, the proportion of misclassified pixels within a certain uncertainty range (e.g., 0.1~0.15) is calculated, and this process is repeated for all samples, resulting in the statistical histogram shown in Figure 11. The color values in the figure represent the total number of samples within the uncertainty interval and pixel error rate interval. The red curve represents the average sample error rate within that uncertainty interval, displayed as the error rate to pixel-level uncertainty curve. The results indicate a distinct sharp rise within the area circled in green. Under the MC Dropout method, samples are mainly concentrated in areas of very low and very high uncertainty, leading to the observed sharp-rise behavior. In contrast, the results with TTA show a smoother variation, with relatively fewer samples having high error rates. Additionally, TTA exhibits high uncertainty predictions for erroneous samples.

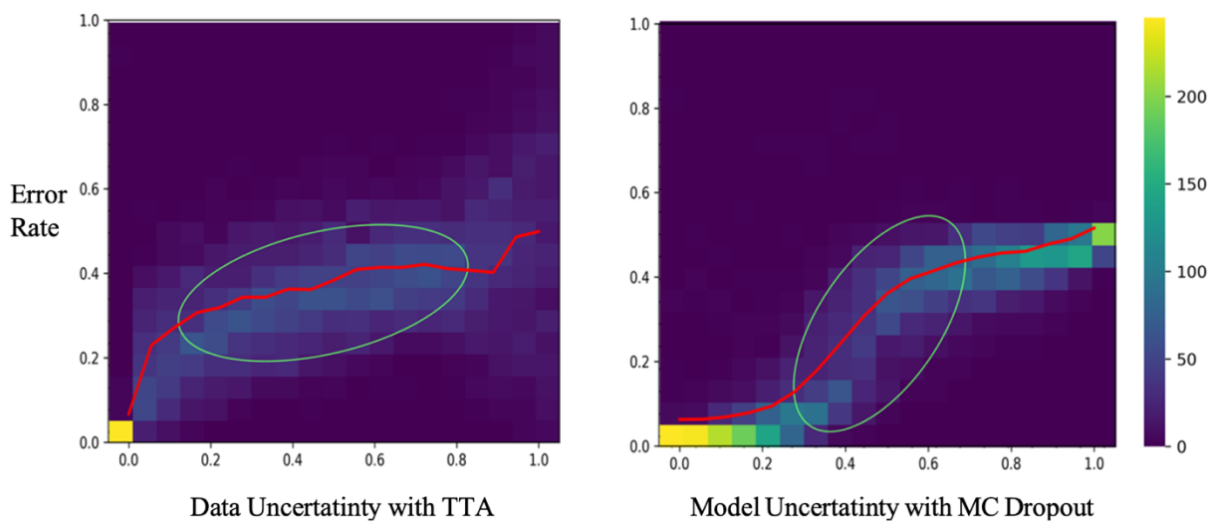

Figure 11: Normalized Statistical Histogram of Predicted Uncertainty-Error Rate

The quantified model uncertainty and data uncertainty are showed in Figure 12(TTA) and Figure 13 ( MC Dropout). The results show that the data uncertainty modeled by TTA can exhibit relatively high uncertainty in areas with segmentation errors, while the model uncertainty modeled by MC Dropout is mainly concentrated at the boundaries of the segmentation results and tends to output lower uncertainty in areas with inaccurate segmentations.

Considering the uncertainty quantification results for both the head and femur data, this study found that TTA's modeling of data uncertainty has good interpretability for segmentation results. It can better recognize erroneous segmentations in non-boundary pixels and can reflect measurement errors made by the model via higher uncertainty score. In a clinical context, doctors can use their own expertise in combination with AI predictions and AI uncertainty quantification results to make informed decisions about whether to accept the measurements by the AI model.

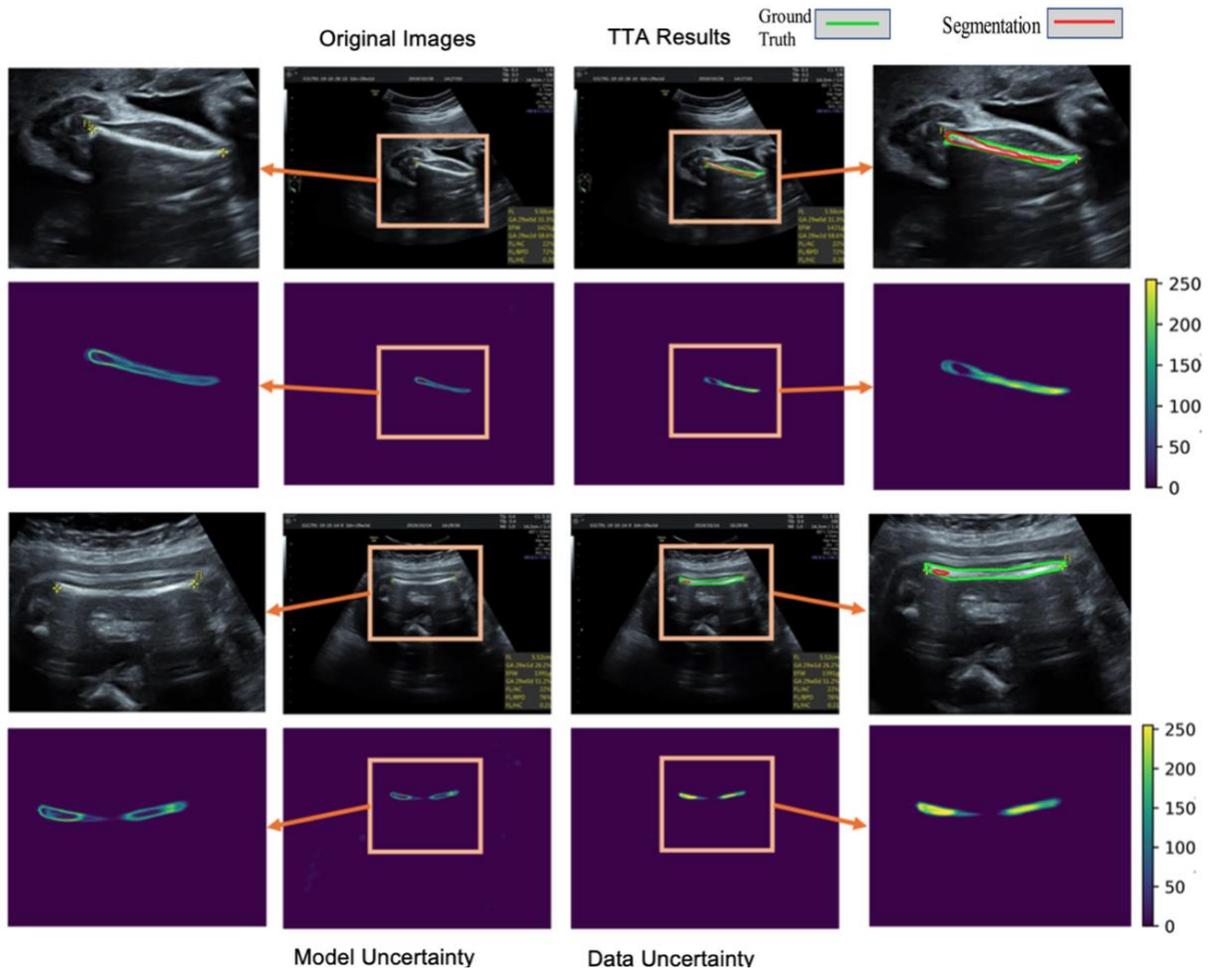

Figure 12. Quantitative Results of Two Types of Uncertainty in Fetal Femur under the TTA Method

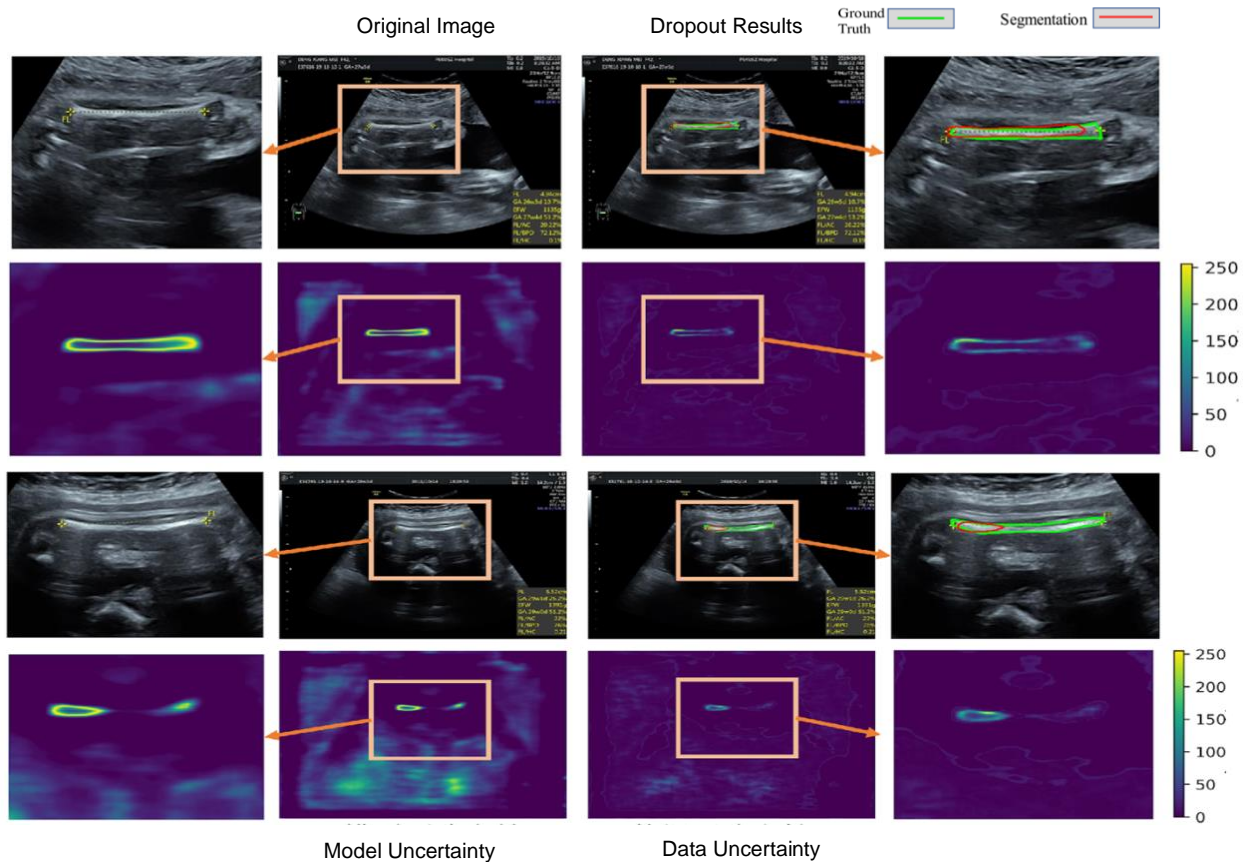

Figure 13. Quantitative Results of Two Types of Uncertainty in Fetal Femur under the Dropout Method

This paper provides uncertainty maps along with the model's predicted segmentation contours and measurement results, which helps doctors to better trust the AI assistants. The measurement process flowchart for clinical assistance in obtaining measurement information is shown in Figure 14. When faced with results exhibiting significant uncertainty, doctors need to manually measure to obtain reliable measurements. When the uncertainty results fall within an acceptable range, doctors consider the measurements to be reliable and can directly output the AI's measurement results. This process significantly enhances clinical efficiency, streamlines the workflow for doctors, and allows them to focus more on exceptional cases.

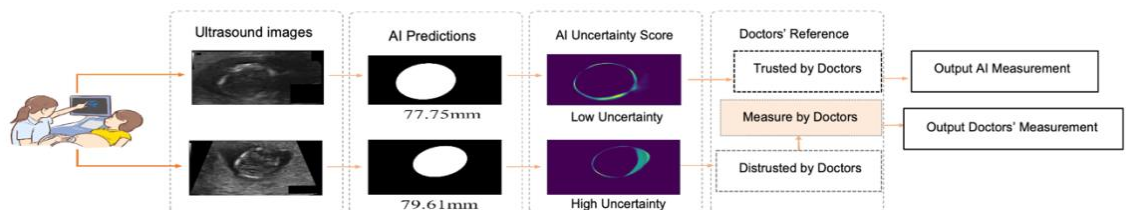

Figure 14. Workflow of the AI-assisted Clinical Parameter Measurement Process with Uncertainty Informed.

## D. Uncertainty Estimation of Out-of-Domain Sample

Uncertainty estimation of out-of-domain(OOD) sample refers to the model's feedback when it is trained on data within the same domain and is subjected to OOD attacks using data with a different distribution from that domain. If the model's uncertainty estimation for OOD data is high, the model exhibits good robustness against adversarial attacks, and its results are more reliable [13]. In this section, the training data in domain is fetal head ultrasound images, while the out-of-domain data is fetal femur ultrasound. The results of uncertainty measurement using TTA data are shown in Figure 15(b) and Figure 15(f). The results indicate that the model tends to have a high uncertainty estimation for OOD data, which can alert doctors to make decisions cautiously.

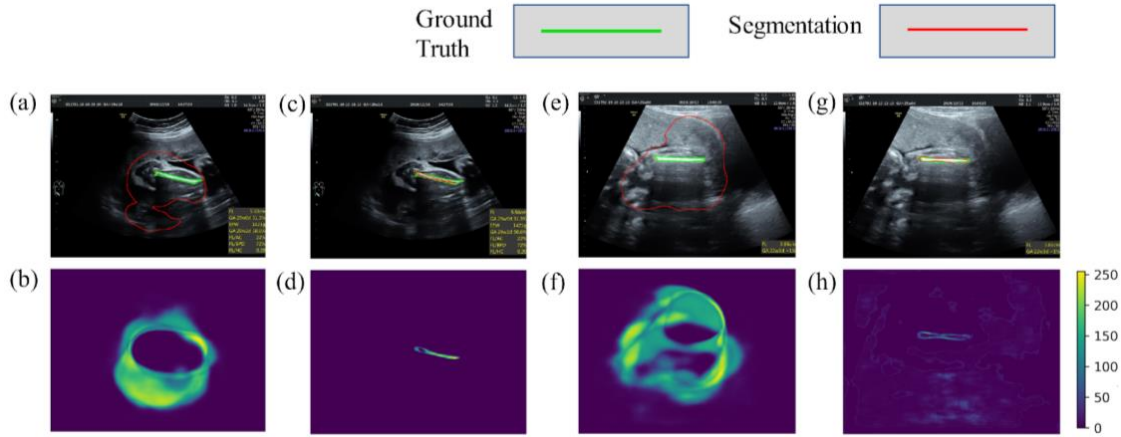

Figure 15. The uncertainty estimation for out-of-domain data. Image (a) and (e) depict the OOD prediction of femur images on the head circumference segmentation model. Image (c) and (g) show the in-domain contour prediction of femur images on the femur segmentation model. Image (b), (d), (f), and (h) present the uncertainty prediction results for TTA data.

## V. Conclusion

Using the standard sections of fetal head and fetal femur ultrasound images, this paper introduces methodologies for extracting target contours and explore techniques for precise parameter measurement. Additionally, uncertainty modeling methods is employed to enhance the training and testing processes of the segmentation network. In the testing phase, TTA and MC Dropout is utilized to assess pixel-wise data uncertainty and model uncertainty on the test set.

The results indicate that the average absolute error in fetal head circumference measurement is 8.0833mm, with an average relative error of 4.7347%. Similarly, the average absolute error in fetal femur measurement is 2.6163mm, with an average relative error of 6.3336%. In the uncertainty modeling experiments, the TTA method demonstrates effective interpretability of data uncertainty on both datasets. This suggests that employing data uncertainty based on the TTA method can support clinical practitioners in making informed decisions and obtaining more reliable measurement results in practical clinical applications.